\documentclass[aip,amsmath,amssymb
preprint,
]{revtex4-1}

\usepackage[english]{babel}

\usepackage[T1]{fontenc}
\usepackage[latin9]{inputenc}
\usepackage{color}

\usepackage{graphicx}
\usepackage{dcolumn}
\usepackage{bm}

\renewcommand{\vec}[1]{\boldsymbol{#1}}

\begin{document}

\preprint{AIP/XXX-XXX}

\title[Magnetic PDMS with aligned Nickel coated CF]{Magnetically Responsive PDMS with aligned nickel coated carbon fibres}

\author{David C. Stanier}
\email{david.stanier@bristol.ac.uk}
\affiliation{Advanced Composites Centre for Innovation and Science, University of Bristol, BS8 1TR, Bristol, UK} 
\author{Jacopo Ciambella}
\email{jacopo.ciambella@uniroma1.it}
\affiliation{Dipartimento di Ingegneria Strutturale e Geotecnica, Sapienza Universit\`a di Roma, 00184, Rome, Italy}
\altaffiliation[Also at ]{Advanced Composites Centre for Innovation and Science,  University of Bristol, BS8 1TR, Bristol, UK}
\author{Sameer S. Rahatekar}
\affiliation{Advanced Composites Centre for Innovation and Science, University of Bristol, BS8 1TR, Bristol, UK} 

\date{\today}

\begin{abstract}

We detail a technique to produce actuators able to bear large strain and respond to an external magnetic field. The material used is PDMS reinforced with nickel coated carbon fibres. Thanks to the nickel functionalisation, the fibre orientation can be achieved by embedding the viscous solution into a low external magnetic field ($<0.2$~T).  It is shown that both mechanical and magnetic properties can be controlled by tailoring the material anisotropy through properly orientating the reinforcing fibres in the pre-curing phase.
The large strain behaviour is investigated by tensile testing up to 60\% of deformation and shows a strong dependence on the fibre orientation. The magnetic properties are investigated by placing beam-like specimens into a uniform magnetic field. The results show a multistable behaviour with a transition from a bending-only deformed configuration for the 0$^\circ$ fibres specimen, to a twisting only configuration, achieved for fibres at 90$^\circ$ whereas all the intermediate angles show both bending and twisting. This behaviour is accurately captured by the large rotations beam model introduced. Such an actuator can be used in all applications which require fast response times and large strain.
\end{abstract}

\pacs{Valid PACS appear here}
\maketitle

\section{Introduction}

To enhance the mechanical properties of elastomer materials, reinforcements are often used. The use of discontinuous fibres is often motivated by the increased surface area and therefore the increased surface interaction with the matrix, when compared to spherical inclusions (such as Carbon black). However they are usually randomly dispersed to give an overall isotropic behaviour, except in those cases when the manufacturing process imparts some anisotropy (e.g. extrusion).\

However, by controlling the orientation of the discontinuous fibres in an elastomeric material, is it possible to affect both the small and large strain behaviour of the material. Conventional composite materials often utilize this orientation with continuous fibres to counter many different problems (e.g. see forward swept divergent wing \cite{KIM12_43}), however additional non-linear effects can be observed when using discontinuous reinforcements at large strain (\textgreater 20\%) \cite{STAN14_95}. This provides additional design space as it is possible to control the reinforcement at discrete locations in 3-dimensions, to affect the material performance at large and small strain.\

The challenges in controlled orientation of materials in solution have been studied \cite{RIKK14_10}, and have included various manufacturing methods, for example, shear \cite{SILV06_27}, electrical \cite{MART05_46} and magnetic alignment \cite{ERB12_335,PROL13_46,CAMP07_45}; the latter of which has demonstrated controlled 3-dimensional reinforcement \cite{ERB12_335}. Significant research has been employed to understand the factors underpinning the orientation of magnetic fibres in a solution, such as field strength \cite{KITA11_18}, field direction \cite{ERB12_335}, and particle morphology \cite{ERB12_335}, in order to spatially control the reinforcements. This has delicate consequences upon the mechanical properties, but can also add a degree of multi-functionality. For instance, the aligned fibres can also provide control of other material properties such as the electrical conductivity \cite{KIM12_49, MART05_46}, dielectric \cite{SHAR10_361}, energy dissipation \cite{CAMP07_45}, gas permeability \cite{SHAR10_361} and thermal conductivity \cite{CHOI03_94}.

It is also possible to utilize the reinforcement and material properties to actuate a specimen of the material. Actuation can be achieved by a variety of different methods, such as pH sensitivity \cite{WU14_15}, swelling \cite{GEET98_39, ERB13_4}, electrical stimulation \cite{BUST08_206} and magnetic attraction \cite{KIMU10_48}. The key advantage of aligning magnetic reinforcements, and using that induced magnetic responsiveness of the material, is that the actuation direction is controlled; both by the non-isotropic mechanical properties and the non-isotropic magnetic responsiveness. 

In this study we take inspiration from nature, and apply some of these concepts to the large strain behaviour of elastomer materials reinforced with discontinuous fibres \cite{ERB12_335}, although the work is equally applicable to any material with a viscous pre-cure phase. Rectangular PDMS specimens with different fibre angles were produced and tested to evaluate the influence of the fibre orientation upon mechanical and magnetic properties. It is shown that the specimens respond to a low magnetic field (<~0.2~T) with large shape variations. Instability effects were also observed and were accurately captured by the large rotations beam model introduced.

Whilst the focus of this investigation has been on the use of a transversely isotropic configuration, an out-of-plane reinforcement \cite{KIMU10_48} or variable angle configuration could be produced by such a method. The use of a nickel coating increases the permittivity of the fibre and therefore allows the use of a relatively low magnetic field that can easily be generated through single or multiple neodymium magnets. The material produced shows the potential for not only improving the properties of elastomer materials in-plane, but also spatially; with consequences upon the mechanical behaviour and multi-functional capabilities.

\section{Materials Preparation}

PDMS (Sylgard 184) is used without further modification. Nickel coated carbon fibres (40\% Ni/C chopped to $\sim$0.25mm, diameter 4.8 $\mu$m) are purchased from Marktek Inc. and used as received. NiC short fibre composites are made by direct mixing of the two compounds; this involves the addition of a carefully weighed amount of NiC fibres to a test tube of $\sim$24g PDMS. This mixture is firstly agitated to partially homogenise the mixture, then sonicated for 1 hour to further separate agglomerated fibres, and then again agitated. The mixture is then left for 1 hour so that any remaining agglomerates settle to the bottom; this assures a homogeneous dispersion of fibres upon curing. The supernatant is then separated and added to curing dishes in 1g quantities and curing agent is added at a ratio of 1:10, with the remaining PDMS residue and NiC sediment cured.

\begin{figure}[htp]
  \begin{center}
  \includegraphics[width=16cm]{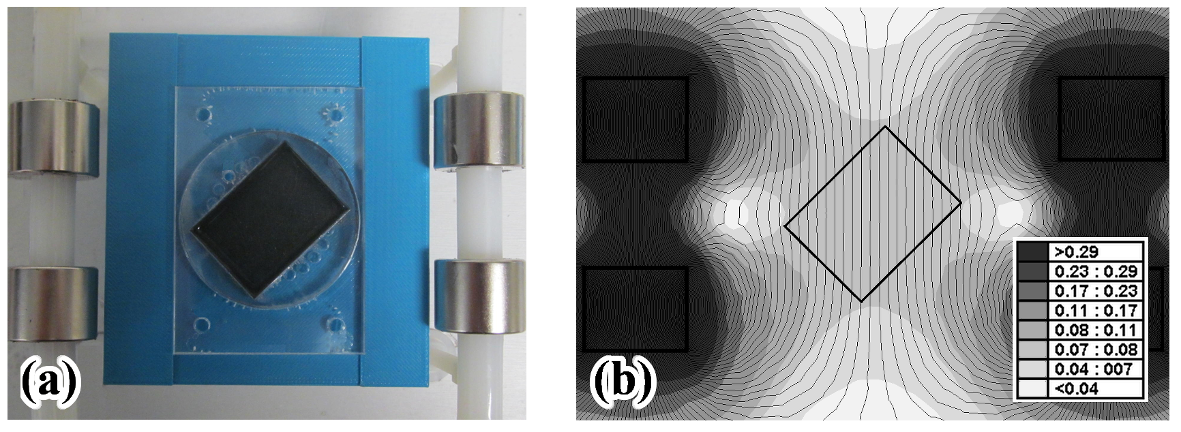}
  \end{center}
\caption{(a) Magnetic setup for specimen curing. The specimen is suspended in the centre of four Neodymium N52 magnets in a homogeneous magnetic field. (b) The magnetic field intensity calculated in the FE environment FEMM \cite{Meeker2006}.}
\label{f:MagneticSetupa}
\end{figure}

Each curing dish is then positioned between four Neodymium N52 magnets in an approximate homogeneous magnetic field of $\sim$0.075 Tesla (see Fig.~\ref{f:MagneticSetupa}). The mixture is then allowed to cure at room temperature for 48 hours. The magnetic setup is assembled from plastic, and arranged so that an approximately homogeneous magnetic field is present on the curing material. Preliminary tests are performed on the arrangement to ensure the homogeneity of the fibres post-cure, and that orientation of the fibres has occurred. The homogeneity of the magnetic field is of importance so that the fibres orient themselves parallel to the magnetic field due to the magnetic torque; if the magnetic field was non-homogeneous then the fibres would also translate and hence agglomerate close to the boundary of the petri dish. As such the resulting material would be non-homogeneous. 

\vspace{-0.3cm}

\begin{figure}[htp]
  \begin{center}
\includegraphics[width=9cm]{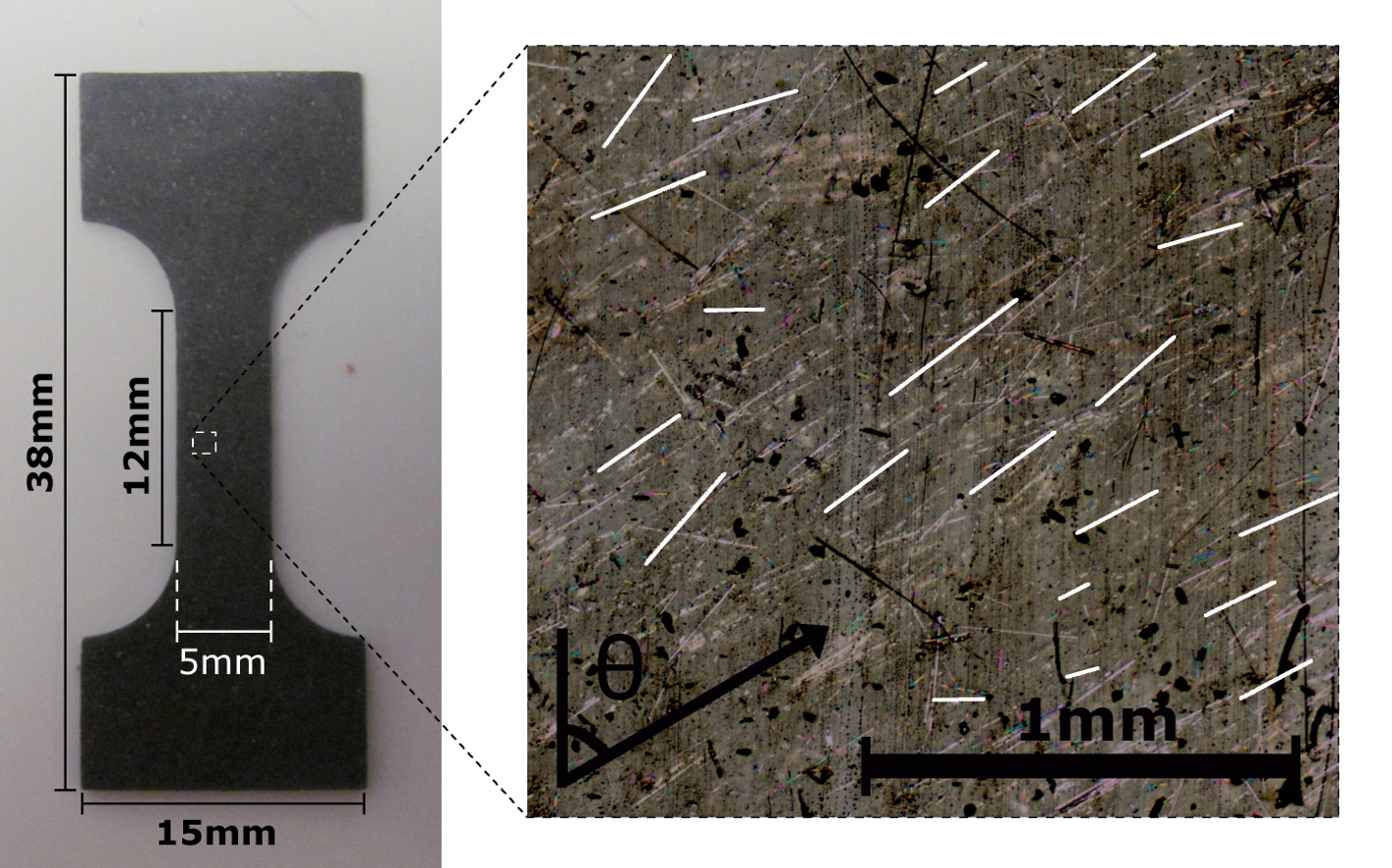}
  \end{center}
\caption{Dumbbell test coupon and inset micrograph of a 60$^{\circ}$ specimen. Selected fibres highlighted in white.}
\label{f:MagneticSetupb}
\end{figure}

Once cured, the specimens are cut from the petri dish at predefined fibre orientation angles between 0$^{\circ}$ and 90$^{\circ}$ degrees, 2 specimens cut from each petri dish with a dumbbell cutting tool (Fig.~\ref{f:MagneticSetupb}). A total of 6 specimens are made at each angle $\theta_0= $[0, 15, 30, 45, 60, 75, 90]. 
A concentration of NiC of 6.0 wt\% is calculated post-cure. This is calculated by burn-off of the PDMS mixture. The residue is added to a crucible and heated to 500$^{\circ}$C for 6 hours, the remnants of this burn-off are then weighed and then compared to that of a pure NR burn-off. This gives the amount of NiC not used in the specimens; with the total amount of NiC known, the amount used in each specimen can be calculated. 

Prior to mechanical testing, an image of each specimen is captured by optical microscopy (see Fig.~\ref{f:MagneticSetupb}). This micrograph of the specimen is post-processed so that a 60 mm x 40 mm section of the specimen is analysed to determine a representative set of data regarding the position, orientation and size of the particles in each specimen. Image software is used to create an image layer onto which the the particles are drawn on; this layer is then imported to the matlab Image Processing Toolbox. 
 
\begin{figure}[htp]
  \begin{center}
\includegraphics[width=7cm]{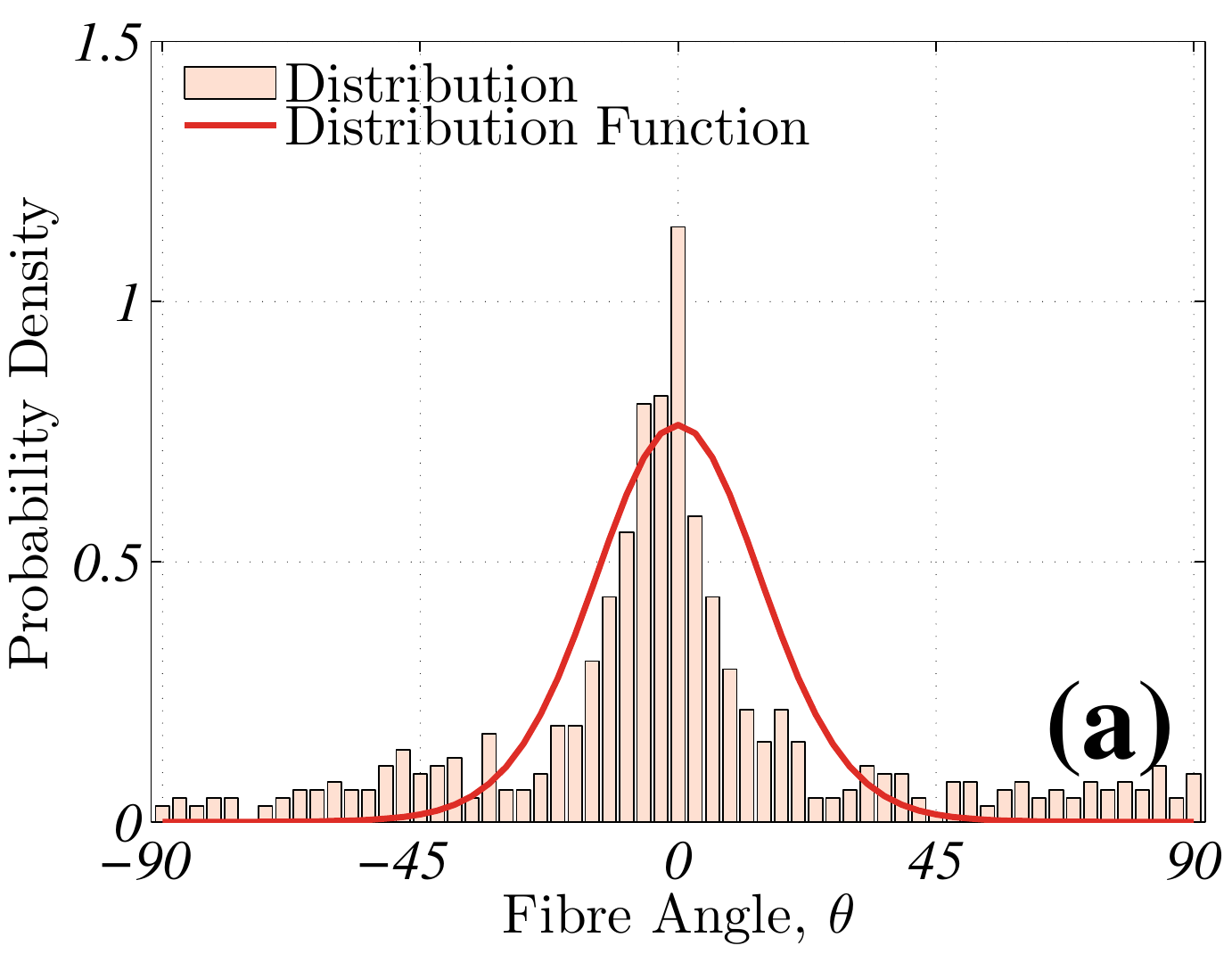}\\
\includegraphics[width=7cm]{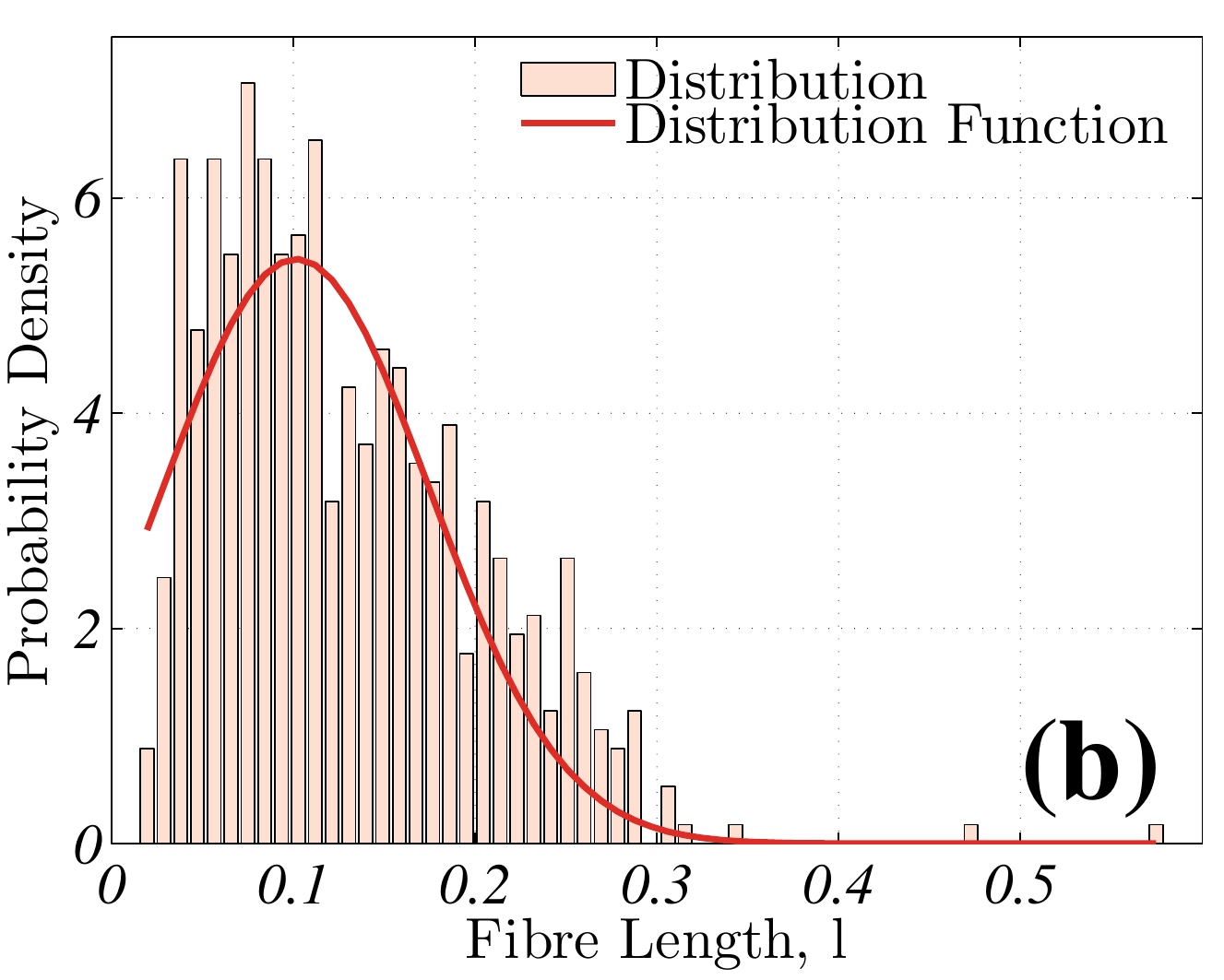}
  \end{center}
\caption{Distributions extracted from a micrograph of a $0^{\circ}$ Specimen, sampling >600 fibres. Average parameter values shown with standard deviations for 6 samples. (a) Fibre Orientation Distribution, $\rho(\theta) =  \exp(b \;cos(2\theta))/\left(2 \int_{0}^{\pi} \exp(b \;cos(2\theta)) d\theta\right)$ (b = 4.49 $\pm$0.57). (b) Fibre Length Distribution, $h(L) = \frac{1}{(\sqrt{2\pi}\; c)}\exp(\frac{-(L-d)^2}{(2c^2)})$ (c = 0.068 $\pm$0.011, d = 0.089 $\pm$0.017).}
\label{f:Distribution}
\end{figure}

Figure ~\ref{f:Distribution}a \& b show the typical distribution curves for a specimen. The orientation shows a significant preferential direction aligned to the direction of the magnetic field, although there is an underlying homogeneous dispersion of fibres that do not orient. Presumably the nickel coating is no longer present in these cases. There is also some distribution in the length of the specimen, which may have some contributing effects upon the ability to orientate in a magnetic field (see Reference\cite{ERB12_335}). 

\section{Experimental results and discussion}
\label{Results}

%
\subsection{Mechanical Properties}

All mechanical tests were performed at room temperature, with dumbbell specimens conforming to ASTM standard D1708, with a specimen thickness of approximately 0.5 mm. The nominal stress is determined as the ratio of the measured force to the original specimen cross-sectional area, whilst the strain is the ratio of the extension to the original distance and is calculated by digital image correlation (DIC). Uniaxial mechanical testing is performed using an Instron testing machine. The specimens are tested applying displacement control (10 mm/min) until rupture. 6 specimens at each angle were tested. 

\begin{figure}
\centerline{
\includegraphics[width=8cm]{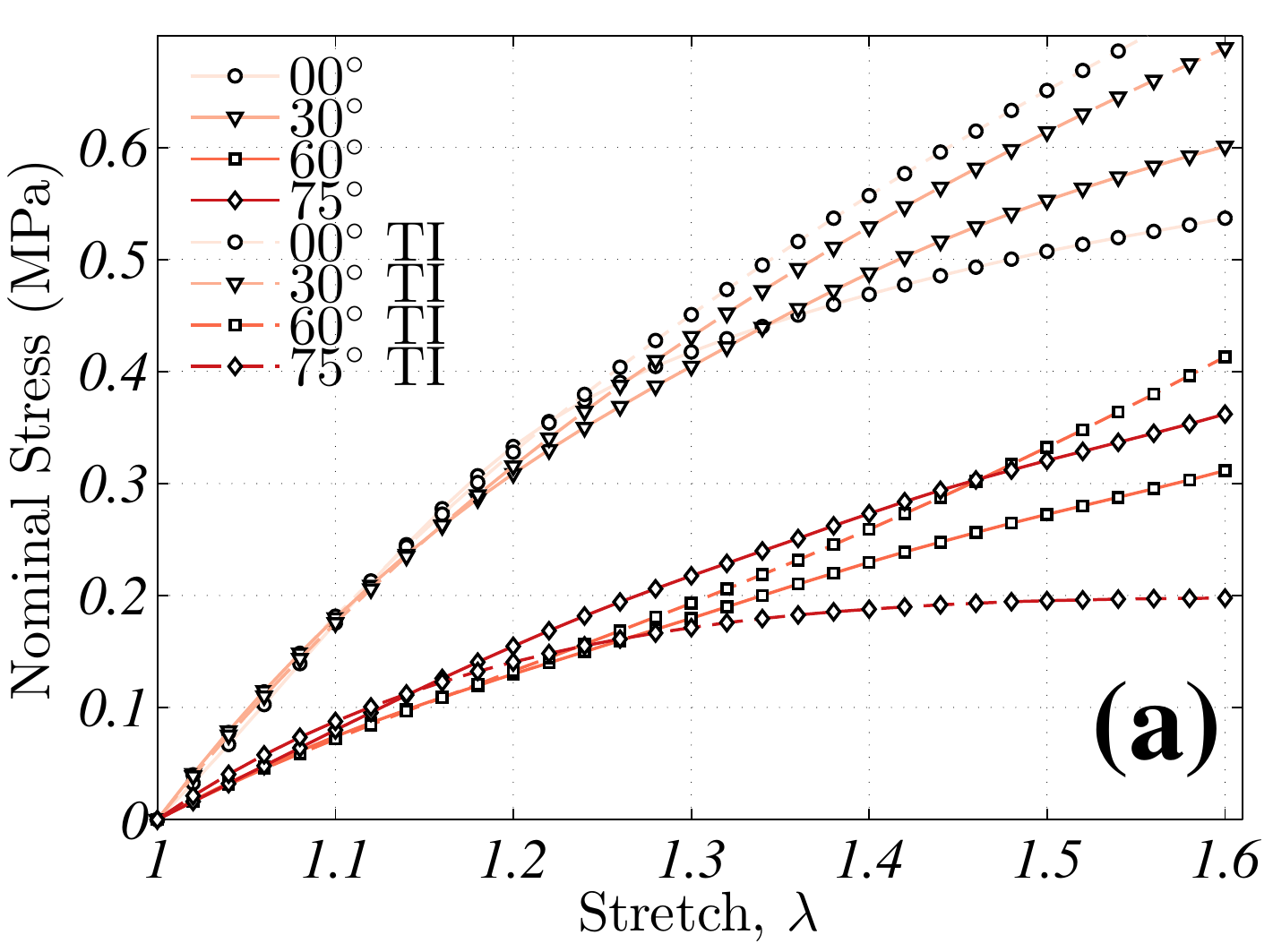}
\includegraphics[width=8cm]{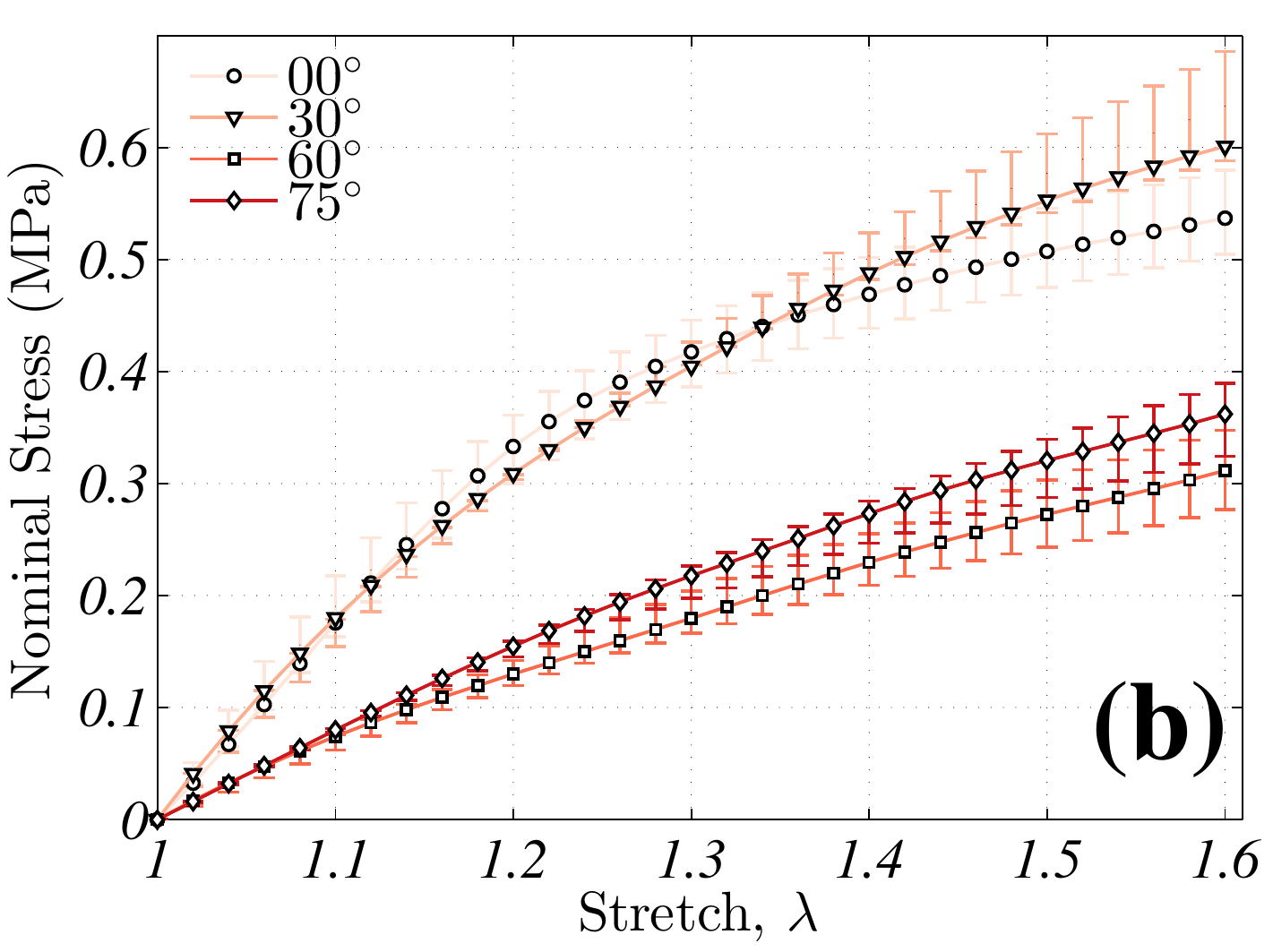}
}
\caption{(a) Large strain experimental results of four specimens compared to the transversely isotropic (TI) constitutive model derived from Eq.~\ref{TransverseIsotropy}. (b) Stress-strain curves, showing confidence intervals of 90\% derived from the stress-strain data of each angle.}
\label{f:LargeStrain}
\end{figure}

The results of the large strain tensile tests are shown in Fig.~\ref{f:LargeStrain} for $0^\circ$, $30^\circ$, $60^\circ$ and $75^\circ$. The differences in the stress-strain curves highlight the different stress transfer mechanisms that take place during deformation and cause the large variation of the tangent modulus seen in the figure. When the fibres are oriented in the loading direction ($0^\circ$ and $30^\circ$) they are able to carry the majority of the load and a high modulus is achieved; on the contrary low moduli correspond to configurations where the fibres are almost orthogonal to the loading ($60^\circ$ and $75^\circ$). 

\begin{figure}
\centerline{
\includegraphics[width=8cm]{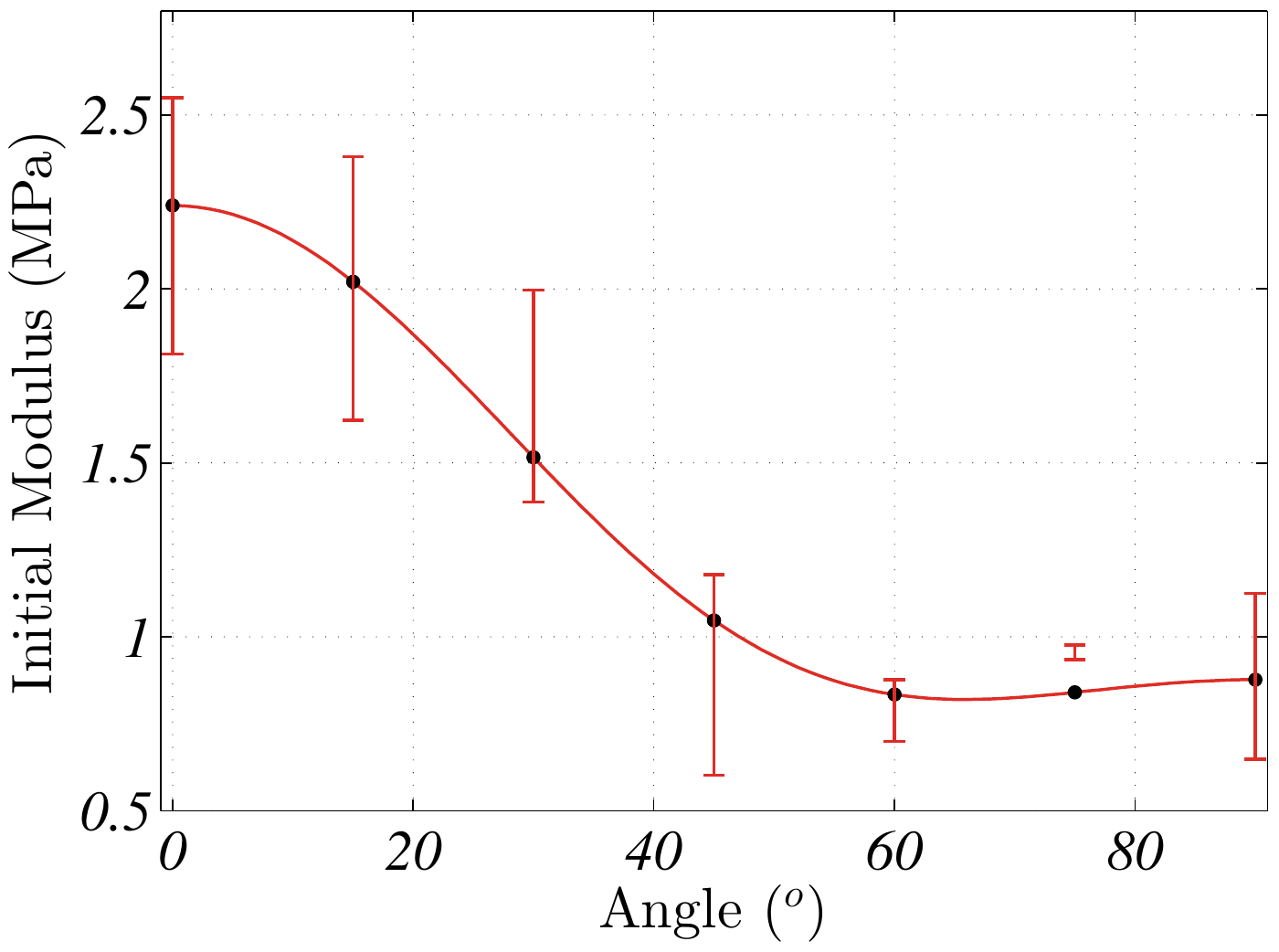}
}
\caption{Variation of the Young's modulus in terms of fibre orientations fitted to Eq.~\ref{CompositeModulus} ($\mu$~=~1.579$\times 10^{-1}$ MPa, $\alpha$ = 1.850, $\beta$ = 2.878). Confidence intervals of 90\% shown for each angle.}
\label{f:SmallStrain}
\end{figure}

It also is interesting to observe the effects of the fibre rotation on this stress transfer mechanism, which can be evidenced when comparing the reinforcement at $0^\circ$ and $30^\circ$ in Fig.~\ref{f:LargeStrain}b. The initial modulus of $0^\circ$ is higher, but at $\lambda$ = $\sim$1.35 the rotation of the fibre towards the loading direction causes the $30^\circ$ tangent modulus to decrease less and the stress to rise above the other. A similar occurrence is not observed between the $60^\circ$ and $75^\circ$ specimens, however this is most likely due to the relatively moderate change in modulus that a rotation of the fibres would cause at these angles (see Fig.~\ref{f:SmallStrain}).

These results can be effectively interpreted by comparison with the transversally isotropic hyperelastic model used in Reference\cite{CIAM14_conf}. The main underlying hypothesis of the model was that the strain energy density of the reinforced material is separable into two contributions: one accounting for the overall stretch of the PDMS matrix and the other weighing the stretch along the stiff fibres. This model is based on the assumption that the strain energy stored in the fibres depends only on the fibres elongation \cite{GUO07_38, CIAM14_conf}. The corresponding nominal stress (force over reference area) in the loading direction is given by

\begin{equation}
S(\lambda) = \Psi_4 \,\lambda^3\cos^2(\theta_0)- \Psi_2\,\lambda^3 + \Psi_1 \,\lambda^3 -p\,\lambda\,,
\label{TransverseIsotropy}
\end{equation}
with $\theta_0$ being the initial orientation of the fibres, $\lambda:=\ell/\ell_0$ the stretch and $\Psi_1$, $\Psi_2$ and $\Psi_4$ constitutive functions defined as
\begin{align}
\Psi_1=\mu\,,\qquad\Psi_2=\alpha\,\mu\,\qquad 
\Psi_4=\beta  \,\mu  \,\left(1-\frac{\lambda^{3/2}}{\left(\lambda ^3 \cos ^2(\theta_0)+\sin ^2(\theta_0)\right)^{3/2}}\right)
\end{align}

The unknown pressure field $p$ in Eq.~\eqref{TransverseIsotropy} can be worked out by enforcing the plane stress condition (null stress in the through-thickness direction). Note that the constitutive parameter $\beta$ weighs the anisotropy of the specimen and in fact the isotropic Mooney-Rivlin model is recovered if $\beta=0$.

For such a model, the expression of the effective tangent modulus (Young's modulus) in the direction of the tensile loading is
\begin{align}
E(\theta_0)= 
\frac{3 \mu}{16} \left[16+ 16\alpha + 5\beta + 8 \beta \cos \left(2 \theta _0\right) + 3 \beta \cos \left(4 \theta _0\right) \right].
\label{CompositeModulus}
\end{align}
Equation~\ref{CompositeModulus} is able to describe the dependence of the Young's modulus on the fibre orientation well, as shown by the fitting in Fig.~\ref{f:SmallStrain}. $E(\theta)$ has a minimum for fibres oriented at $\theta_0=65.9^\circ$ as corroborated by the stress-strain data in Fig.~\ref{f:LargeStrain}. For this angle the fibre provides the least amount of resistance to the uniaxial forces imposed upon the specimen. On the contrary for $0^\circ$ or $90^\circ$, the effective modulus has a local maxima (see Fig.~\ref{f:SmallStrain}). The maximum at $0^\circ$ being significantly larger than at $90^\circ$. However, the limiting value at $90^\circ$ is interesting as it indicates the potential importance of the lateral constraint of the reinforcement in an incompressible system. 

At strains above $\sim$30\% the model begins to struggle to capture the highly non-linear behaviour observed in the experimental data in Fig.~\ref{f:LargeStrain}a. Therefore a more complex form of the constitutive parameters $\Psi_1$, $\Psi_2$ and $\Psi_4$ may be considered but it would be out of the scope of this paper; in fact, as shown in the next section, the deformations obtained by magnetically actuating a thin specimen are rarely larger than $5$\% making this model adequate to describe the material behaviour. 

\begin{figure}[htp]
  \begin{center}
\includegraphics[scale=1]{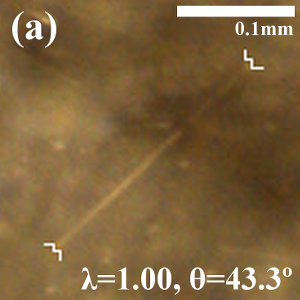}
\includegraphics[scale=1]{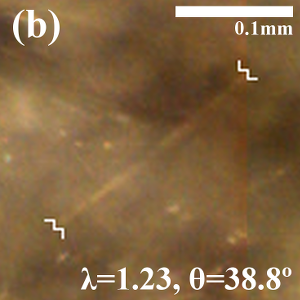}
\includegraphics[scale=1]{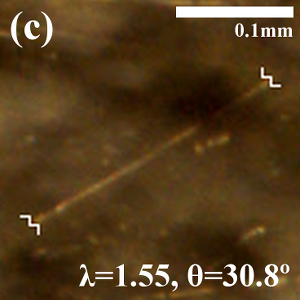}
  \end{center}
  \caption{Experimental micrographs of fibre rotation in the PDMS matrix at different stretch levels, for a specimen with initial angle of the fibres at 45$^{\circ}$.}
  \label{f:ExpFibreOrientation}
\end{figure}

\begin{figure}
\centerline{
\includegraphics[width=8cm]{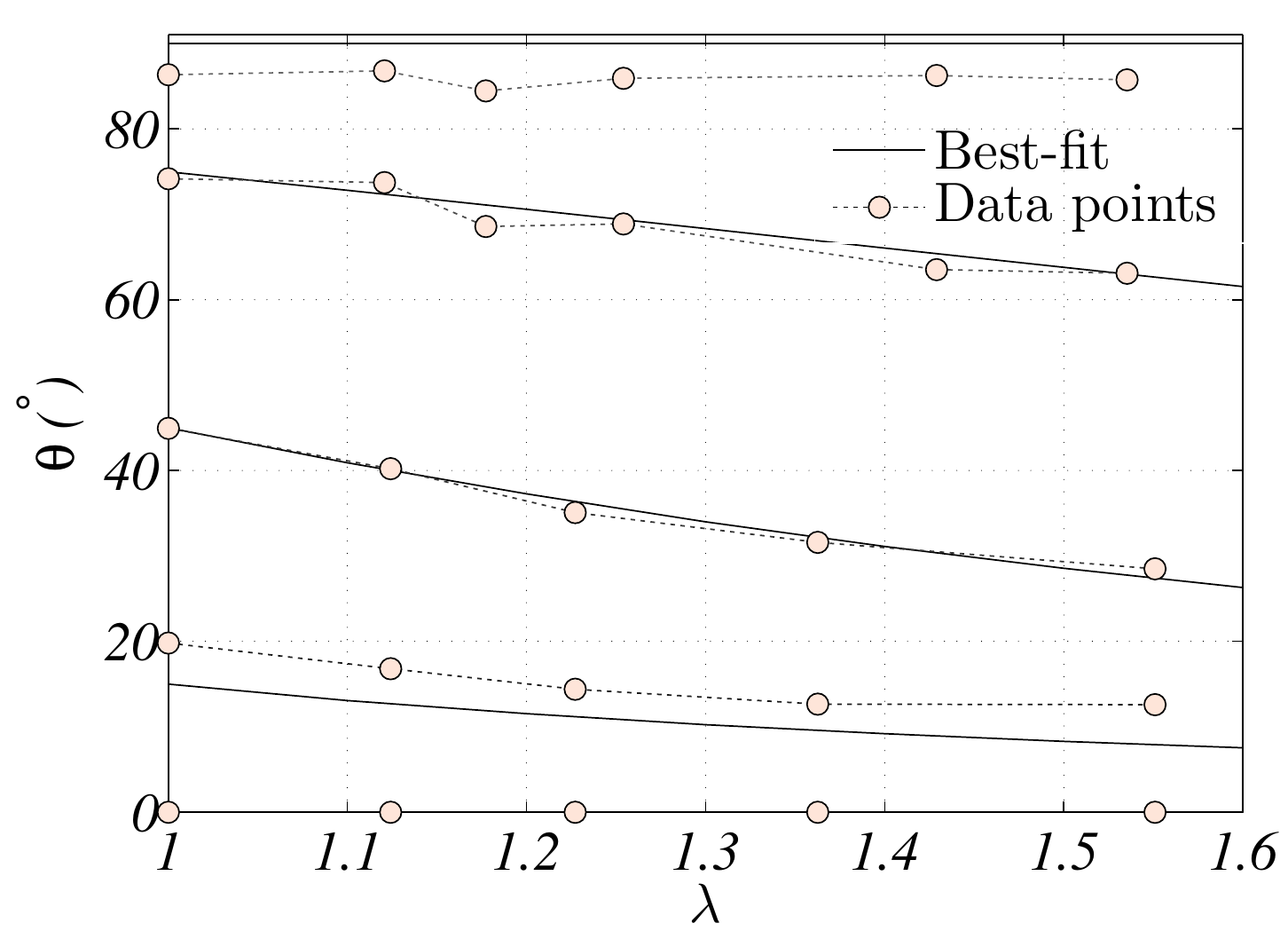}
}
\caption{Rotation of individual fibres compared to the predicted rotation of the constitutive model.}
\label{f:FibreOrientation}
\end{figure}

To further analyse the transversally isotropic nature of the produced composite materials, during tensile testing, microscopy is undertaken on identical specimens at strains values up to $\sim$55\% (see Fig.~\ref{f:ExpFibreOrientation}). This has allowed the rotation of individual fibres to be captured and compared to rotations predicted by the constitutively transversely isotropic model. By using Nanson's formula (see, e.g., Ref.\cite{OGDE97_book}), it is possible to obtain the current orientation of the fibre $\theta$ in terms of the original angle $\theta_0$ and the stretch $\lambda$, i.e., $\theta = -\tan^{-1}\left[\lambda^{-3/2} \tan(\theta_0)\right]$. 
The results of the fitting with this formula are shown in Fig.~\ref{f:FibreOrientation} for five specimens with initial angles $0^\circ$, $15^\circ$, $45^\circ$, $75^\circ$ and $90^\circ$. A remarkably close agreement between the rotation predicted by the model and the actual rotation of the fibres is apparent up to $55\%$ of deformation, which suggest the effectiveness of the plane stress assumption applied in Eq.~\eqref{TransverseIsotropy}.

\subsection{Magnetic Actuation}

\begin{figure}
\centerline{
\includegraphics[width=8cm]{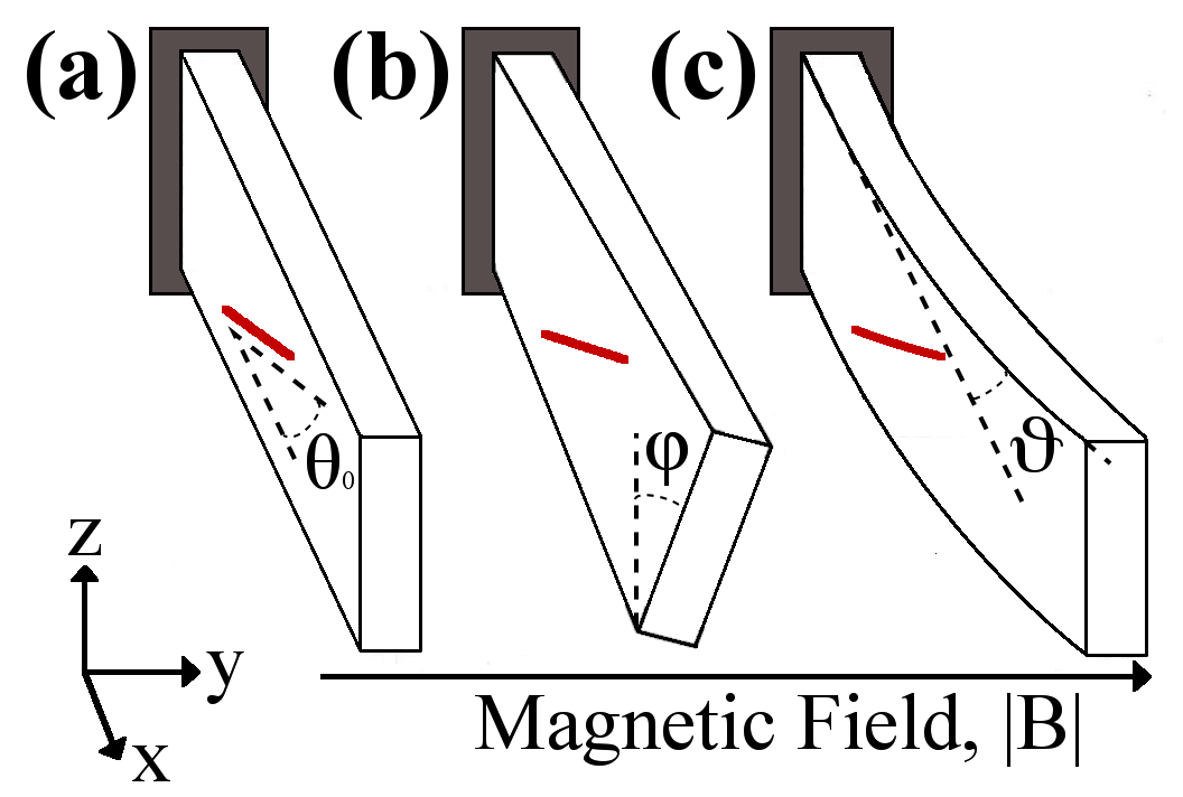}
}
\caption{Schematic representation of the beam model. (a) The undeformed beam showing the reinforcement angle in the x-z plane, $\theta_0$. (b) The twisting angle, $\varphi$. (c) The bending angle, $\vartheta$. }
\label{f:BeamModel}
\end{figure}

The actuation capabilities of the material were tested for each fibre orientation by cutting a rectangular shaped specimen and suspending it between the plates of an electromagnet in a cantilever configuration (as sketched in Fig.~\ref{f:BeamModel}). The resulting magnetic field is homogeneous and its intensity can be controlled by varying the current absorbed by the electromagnet. Further to this, the dynamic response of the actuation was viewed by generating a rotating magnetic field at frequencies between 0 and 3Hz through cylindrical neodymium magnets; demonstrating its capabilities for fast response times.

The advantages of having the fibres in a plane orthogonal to the magnetic field would be the possibility of having different actuator responses based on the fibre angle $\theta_0$. 
To investigate this behaviour we introduce a large rotations beam model for which the total energy is the sum of an elastic and a magnetic part, i.e., 
\begin{equation}
E_{tot}=h\,t\int_0^L (\Psi_e+\Psi_m)\;\text{d}s
\label{Total Energy}
\end{equation}
where $h$ and $t$ are the beam width and thickness respectively, and $\Psi_e$ and $\Psi_m$ given by
\begin{equation}
\Psi_e  = \frac{\text{EI}}{2\,L^2}\;\vartheta '(s)^2+\frac{G J}{2\, L^2}\;\varphi '(s)^2\,,\qquad \Psi_m =-\chi (\vec{n}.\vec{B})^2\,.
\label{Energy_Densities}
\end{equation}
Here the elastic energy is expressed in terms of the bending, $\vartheta$, and twisting, $\varphi$, angles and accounts for large rotations/displacements; $\Psi_m$ depends upon the relative angle between the magnetic field $\vec{B}=\lbrace 0,B_y,0\rbrace$ (here assumed to be directed along the y-axis) and the fibre orientation in the deformed configuration $\vec{n}=\{\sin (\theta_0) \sin (\vartheta) \sin (\varphi )+\cos (\theta_0) \cos (\vartheta ),\cos (\theta_0) \sin (\vartheta )-\sin (\theta_0) \cos (\vartheta ) \sin (\varphi ),\sin (\theta_0) \cos (\varphi )\}$ (see Fig.~\ref{f:BeamModel}). In \eqref{Total Energy} $s$ is the abscissa along the specimen length ($0\leq s \leq L$) and $\chi=\chi_a\,\nu_{f}/(2\mu_0)$ where $\chi_a$ represents the magnetic anisotropic susceptibility, $\mu_0$ the vacuum permittivity and $\nu_f$ is the volume fraction of the fibres\cite{KIMU10_48}. $EI$ and $GJ$ are the bending and twisting rigidities, respectively.

The deformed configuration of the beam specimen is studied by looking at the minima of the total energy; the model~\eqref{Total Energy} being nonlinear, only numerical solutions can be derived. However, a great deal of insight into the specimen behaviour can still be gained by taking a fourth order expansion of the total energy with respect to the maximum bending and twisting angles; as a result, one can obtain a closed form approximate expression of the stable and unstable equilibria configurations of the beam. This approximate expression of the total energy in terms of the maximum bending $\vartheta_m$ and twisting $\varphi_m$ angles achieved at the free end of the beam is given by

\begin{align}
&E_{Total} = \frac{1}{15} B_y^2 \vartheta_m^4 L \chi  \cos ^2(\theta_0)-\frac{4}{15} B_y^2 \vartheta_m^3 L \varphi_m \chi  \sin (\theta_0) \cos (\theta_0)+\frac{1}{5} B_y^2 \vartheta_m^2 L \varphi_m^2 \chi  \sin ^2(\theta_0)\notag\\[0.3cm]
&-\frac{1}{3} B_y^2 \vartheta_m^2 L \chi  \cos ^2(\theta_0)-\frac{1}{15} B_y^2 \vartheta_m L \varphi_m^3 \chi  \sin (\theta_0) \cos (\theta_0)+\frac{2}{3} B_y^2 \vartheta_m L \varphi_m \chi  \sin (\theta_0) \cos (\theta_0)\notag \\[0.3cm]
&+\frac{1}{15} B_y^2 L \varphi_m^4 \chi  \sin ^2(\theta_0)-\frac{1}{3} B_y^2 L \varphi_m^2 \chi  \sin ^2(\theta_0)+\frac{\text{EI} \vartheta_m^2}{2 L^3}+\frac{G J \varphi_m^2}{2 L^3}\;.
\label{4thOrderEnergy}
\end{align}

The stability of the equilibria can be studied by looking at the second derivatives of Eq.~\eqref{4thOrderEnergy}, i.e., the Hessian matrix of the system, which gives the following value of the critical magnetic field $B_y^{(crit)}$ 
\begin{equation}
B_y^{(crit)}) = \sqrt{\frac{3\; \text{EI}}{2\, L^4 \,\chi  \left(\text{EI}/\text{GJ} \sin ^2(\theta_0)+\cos^2(\theta_0)\right)}}\,.
\label{bycrit}
\end{equation}
Up to the critical value, $B_y^{(crit)}$, the system has only one equilibrium position that corresponds to the undeformed configuration; as soon as $B_y\geq B_y^{(crit)}$ the initial configuration becomes unstable (local maximum of the energy) and two symmetric minima appear. In this situation the reinforcing fibres attempt to rotate parallel to the magnetic field. The presence of two symmetric minima is due to the fact that positive or negative bending/twisting angles have the same weights in the energy. 

\begin{figure}
\centerline{
\includegraphics[width=8cm]{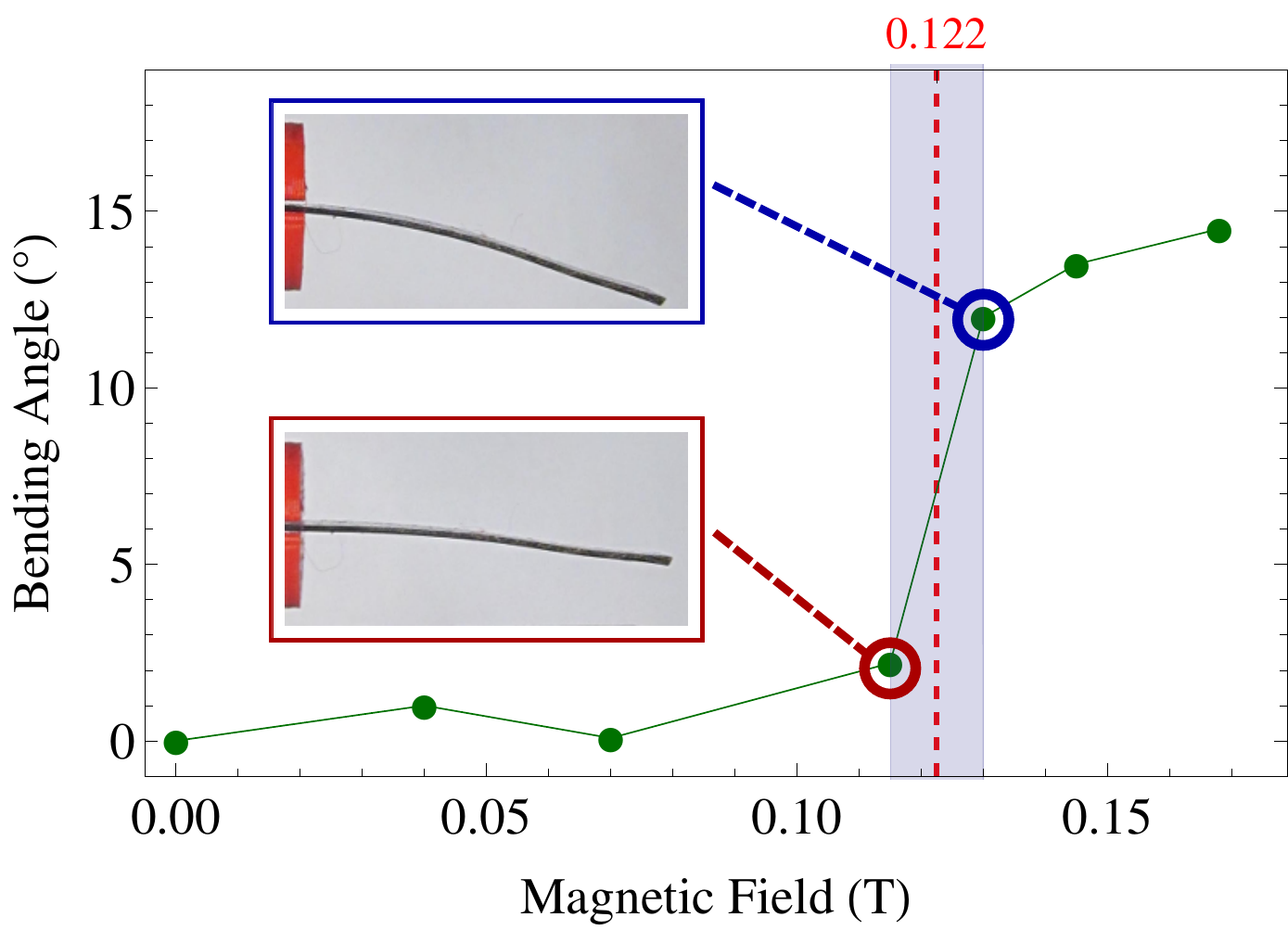}
}
\caption{Maximum bending angle in terms of the intensity of the magnetic field for a specimen with fibres at $\theta_0=30^\circ$. When the magnetic field overcomes $\sim$0.122 T a sudden increase in the bending angle is observed.
}
\label{f:instab30}
\end{figure}

\begin{figure}
\centerline{
\includegraphics[width=8cm]{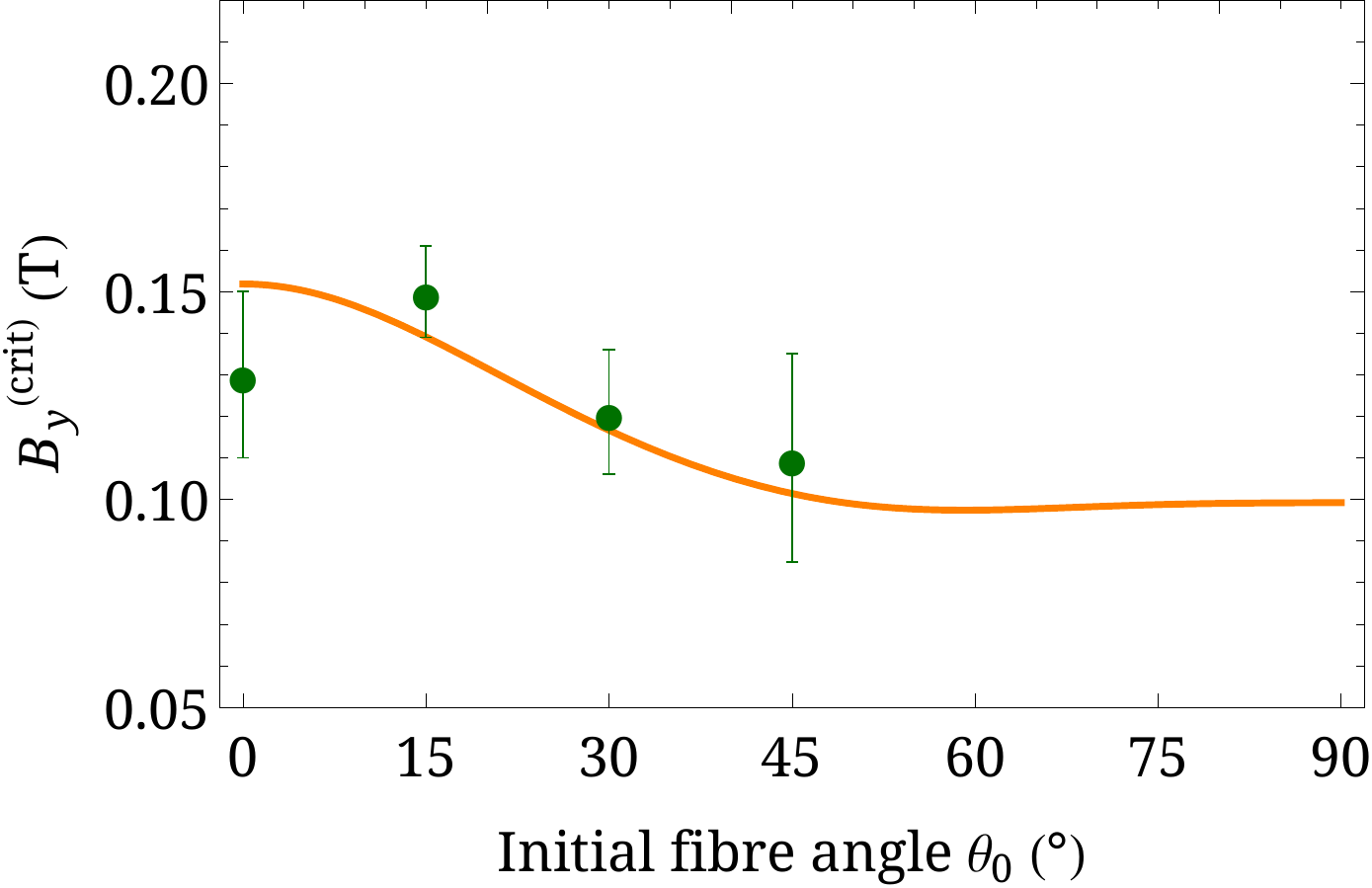}
}
\caption{Critical magnetic field intensity $B_y^{(crit)}$ at which the undeformed configuration becomes unstable, assuming the model~\eqref{bycrit}. (Parameters: E($\theta_0$) and G($\theta_0$) taken from the results of the mechanical testing, I = 7.29$\times 10^{-14}$ (m$^{4}$), J = 2.51$\times 10^{-13}$ (m$^{4}$), L = 25$\times 10^{-3}$ (m), $\chi_a$ = 1.21$\times 10^{-3}$ and $\nu_f$ = 6~\%). The green dots represents the results of the experiment carried out by the authors for specimens with fibres at $\theta_0=\lbrace 0^\circ, 15^\circ, 30^\circ, 45^\circ\rbrace$.}
\label{f:bycrit}
\end{figure}

The value of $B_y^{(crit)}$ can be experimentally determined by slowly increasing the intensity of the magnetic field up to the point at which a sudden jump in the deformation of the beam is observed. This effect is shown in Fig.~\ref{f:instab30} for a specimen with fibres at $\theta_0=30^\circ$ for which it is seen that when the magnetic field overcomes $\sim$0.122 T a sudden increase in the bending angle happens. This is clearly visible from the camera samples in the figure. The values of $B_y^{(crit)}$ predicted by the model through Eq.~\eqref{bycrit} are shown in Fig.~\ref{f:bycrit} against the experimental data for specimens with fibres at $\theta_0=\lbrace 0^\circ, 15^\circ, 30^\circ, 45^\circ\rbrace$. For the other specimens the point of instability is less clear, and a more smooth increase in the bending and twisting angles with the magnetic field intensity is observed. This is partly due to the difficulty in measuring the twisting angle, which becomes the dominant actuation mode for higher angles, which is confused further by the fibres not being completely aligned but having a spread in the orientation as shown in Fig.~\ref{f:Distribution}a. 
The fitting in Fig.~\ref{f:bycrit} was achieved with a value of the anisotropic susceptibility $\chi_a=1.21\times 10^{-3}$; it should be noted that this value is about twenty times larger than the one reported in the literature for neat carbon fibres \citep{KIMU10_48} and is likely due to the nickel functionalisation of the fibre which make them easier to orient in low magnetic fields.

\begin{figure}
\centerline{
\includegraphics[width=8cm]{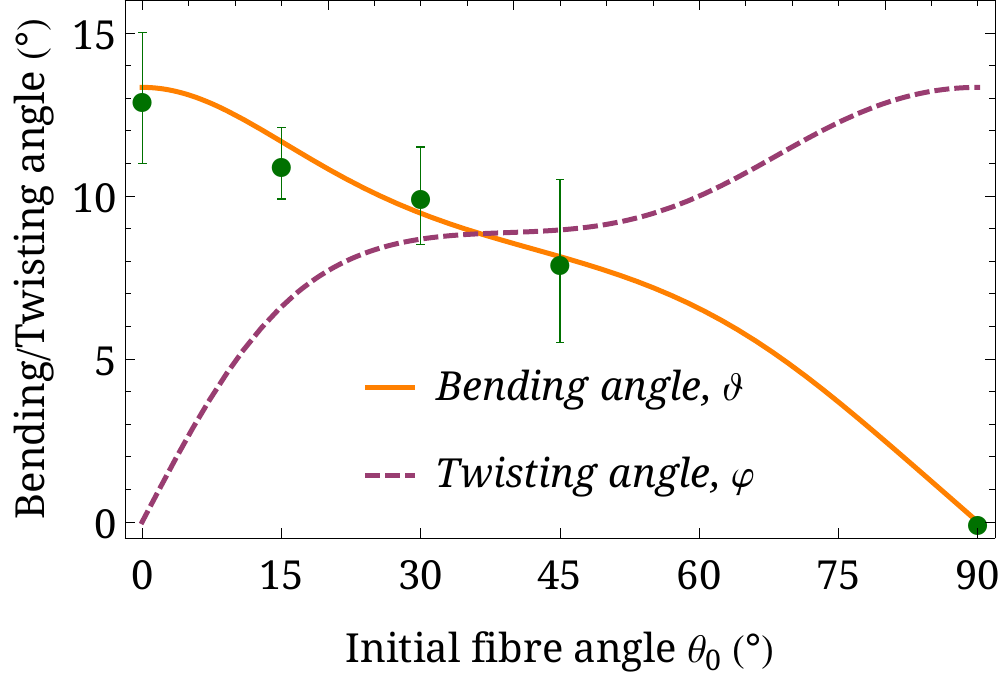}
}
\caption{Maximum twisting and bending angles achieved for an imposed orthogonal magnetic field at the occurrence of instability, assuming the model~\eqref{4thOrderEnergy}. (Parameters: E($\theta_0$) and G($\theta_0$) taken from the results of the mechanical testing, I = 7.29$\times 10^{-14}$ (m$^{4}$), J = 2.51$\times 10^{-13}$ (m$^{4}$), L = 25$\times 10^{-3}$ (m), $\chi_a$ = 1.21$\times 10^{-3}$ and $\nu_f$ = 6~\%. The green dots represents the results of the experiment carried out by the authors for specimens with fibres at $\theta_0=\lbrace 0^\circ, 15^\circ, 30^\circ, 45^\circ, 90^\circ\rbrace$.}
\label{f:ModelMagnetic}
\end{figure}

The minimisation of \eqref{4thOrderEnergy} allows the equilibrium positions for each fibre orientation to be worked out as shown in Fig.~\ref{f:ModelMagnetic}; this latter was obtained with values of E($\theta_0$) and G($\theta_0$) taken from the mechanical characterisation carried out by the authors. At $0^{\circ}$ the specimen exhibits bending only, whereas at $90^{\circ}$ only twisting is observed. Fibre angles in between show both bending and twisting. The results of the experiments are shown in the same figure with green dots representing the bending angles of specimens with $\theta_0=\lbrace 0^\circ, 15^\circ, 30^\circ, 45^\circ, 90^\circ\rbrace$. The corresponding twisting angles are not shown due to the large errors in their estimate; the width of the beam being much smaller than its length.

\begin{figure}[htp]
  \begin{center}
\includegraphics[width=4cm]{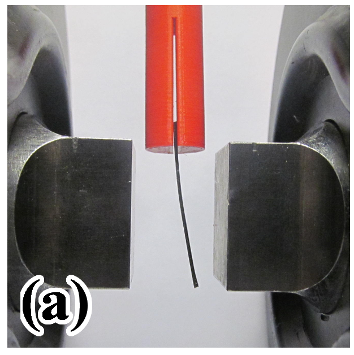}
\includegraphics[width=4cm]{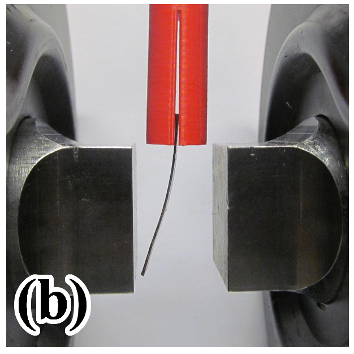}\\
\includegraphics[width=4cm]{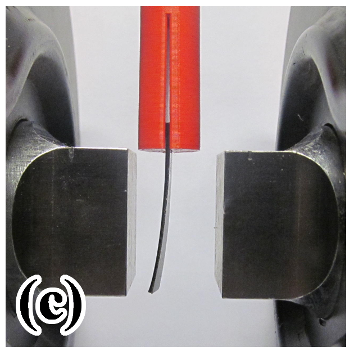}
\includegraphics[width=4cm]{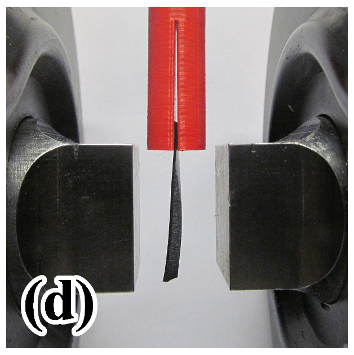}
  \end{center}
\caption{Static Actuation of selected specimens in a homogeneous magnetic field, (a) 0$^{\circ}$, (b) 30$^{\circ}$, (c) 45$^{\circ}$, (d) 90$^{\circ}$. Specimens of length $\sim$25mm, width 7mm and thickness 1mm.}
\label{f:StaticActuation}
\end{figure}

Although a certain degree of dispersion in the fibre angle is present (see the distribution function in Fig.~\ref{f:Distribution}), the actuation of the specimens in the magnetic field confirm the behaviour predicted by the model. Indeed, as seen in Fig.~\ref{f:StaticActuation}, a 0$^{\circ}$ specimen exhibited only bending, whilst only twisting is observed at 90$^{\circ}$. At angles between, either twisting or bending is observed as the dominant actuation mechanism; although it is still possible to observe both (see Fig.~\ref{f:StaticActuation}b \& c). 

\section{CONCLUSIONS}

We have studied the behaviour of magnetically responsive nickel-coated carbon fibres embedded in an PDMS elastomer material and oriented under a magnetic field to produce a transversely isotropic material. 

In this way, the material has mechanical properties that are controlled by the strength and direction of the magnetic field during the cure of the material to produce bespoke material properties. The fibre orientation causes both longitudinal and transverse stiffening effects that are observed both at small and large strains. 

The large strain macroscopic behaviour of the composite is also shown to be described by the transversely isotropic constitutive equation up to around 30\%. In particular, the rotation of the reinforcing fibres is closely matched to the experimental results. Above 30\% strain, the model struggles to capture the non-linear behaviour of the material; likely due to the non-ideal interface between the fibres and the matrix and the consequent debonding of the fibres. 

Thanks to the nickel functionalisation of the fibres, the elastomer material is also responsive to low magnetic fields (<0.2 T) and is shown to actuate due to the reinforcing fibres attempting to align themselves parallel to the magnetic field lines. A beam-like configuration  with fibres aligned within the plane of the beam has been studied. The resulting actuator shows a multistable behaviour for which, up to a critical value of the magnetic field, the undeformed configuration is stable and then becomes unstable for larger magnetic field intensities. Both twisting and bending behaviour is independently observed at different initial orientations of the fibre and predicted by the simple model introduced. 

The result of using magnetically responsive reinforcing fibres is that both mechanical and magnetic responses can be tailored, in a complimentary manner. This could potentially be useful when considering the significant design requirements of a micro-swimmer \cite{GAO14_14}, requiring a remotely actuated flexible material that still has significant stiffness to propel itself through a viscous fluid. The proper control of the direction of the external magnetic field could be used to accurately control the shape of the actuator in a robust manner which is a peculiarity of the proposed configuration.


\bibliographystyle{chicaco}
\bibliography{ECCMRIXV2}

\end{document}